\documentclass[lettersize,journal]{IEEEtran}
\usepackage{amsmath,amsfonts}
\usepackage{algorithmic}
\usepackage{algorithm}
\usepackage{array}
\usepackage[caption=false,font=normalsize,labelfont=sf,textfont=sf]{subfig}
\usepackage{textcomp}
\usepackage{stfloats}
\usepackage{url}
\usepackage{verbatim}
\usepackage{graphicx}
\usepackage{cite}
\usepackage{stfloats}
\usepackage{cite}
\usepackage{bbding}
\usepackage{booktabs}
\usepackage{wrapfig}
\usepackage{graphicx}
\usepackage{multirow}
\usepackage{enumitem}
\usepackage{amsthm}
\usepackage{amsfonts,amssymb}
\usepackage{arydshln}
\usepackage{pifont}
\usepackage{amsmath}
\usepackage{makecell}
\usepackage{url}
\usepackage[switch]{lineno}
\usepackage{amsfonts,amssymb}
\usepackage[justification=centering]{caption}

\begin{document}
\title{A Survey on Federated Recommendation Systems}

\author{
\IEEEauthorblockN{
Zehua Sun\IEEEauthorrefmark{1},  
Yonghui Xu\IEEEauthorrefmark{1}, 
Yong Liu, 
Wei He, 
Lanju Kong,
Fangzhao Wu, 
Yali Jiang\IEEEauthorrefmark{2},
Lizhen Cui
\thanks{Zehua Sun, Yonghui Xu, Wei He, Lanju Kong, Yali Jiang and Lizhen Cui are with Joint SDU-NTU Centre for Artificial Intelligence Research (C-FAIR) \& Software School, Shandong University. Yonghui Xu are also with Sino-Singapore International Joint Research Institute.}
\thanks{Yong Liu are with Alibaba-NTU Singapore Joint Research Institute, Nanyang Technological University, Singapore.}
\thanks{Fangzhao Wu are with Microsoft Research Asia, China.}
\thanks{\IEEEauthorrefmark{1}Zehua Sun and Yonghui Xu are Co-First authors.}
\thanks{\IEEEauthorrefmark{2}Corresponding author: jiang.yl@sdu.edu.cn.}}

}

\markboth{IEEE Transactions on Neural Networks and Learning Systems,~Vol.~14, No.~8, December~2022}%
{Shell \MakeLowercase{\textit{et al.}}: A Sample Article Using IEEEtran.cls for IEEE Journals}


\maketitle

\begin{abstract}
Federated learning has recently been applied to recommendation systems to protect user privacy. In federated learning settings, recommendation systems can train recommendation models by collecting the intermediate parameters instead of the real user data, which greatly enhances user privacy. Besides, federated recommendation systems can cooperate with other data platforms to improve recommendation performance while meeting the regulation and privacy constraints. However, federated recommendation systems face many new challenges such as privacy, security, heterogeneity and communication costs. While significant research has been conducted in these areas, gaps in the surveying literature still exist.  In this survey, we—(1) summarize some common privacy mechanisms used in federated recommendation systems and discuss the advantages and limitations of each mechanism; (2) review several novel attacks and defenses against security; (3) summarize some approaches to address heterogeneity and communication costs problems; (4) introduce some realistic applications and public benchmark datasets for federated recommendation systems; (5) present some prospective research directions in the future. This survey can guide researchers and practitioners understand the research progress in these areas.
\end{abstract}

\begin{IEEEkeywords}
Recommendation Systems, Federated Learning, Privacy, Security, Heterogeneity, Communication Costs.
\end{IEEEkeywords}

\section{Introduction}
\label{sec:introduction}
\IEEEPARstart{I}{n} recent years, recommendation systems have been widely used to model user interests so as to solve information overload problems in many real-world fields, e.g., e-commerce \cite{sarwar2000analysis}\cite{schafer2001commerce}, news\cite{zheng2018drn}\cite{liu2010personalized} and healthcare\cite{yue2021overview}\cite{kim2014item}. To further improve the recommendation performance, such systems usually collect as much data as possible, including a lot of private information about users, such as user attributes, user behaviors, social relations, and context information.

Although these recommendation systems have achieved remarkable results in accuracy, most of them require a central server to store collected user data, which exists potential privacy leakage risks because user data could be sold to a third party without user consent, or stolen by motivated attackers. In addition, due to privacy concerns and regulatory restrictions, it becomes more difficult to integrate data from other platforms to improve recommendation performance. For example, regulations such as General Data Protection Regulation (GDPR)\cite{albrecht2016gdpr} set strict rules on collecting user data and sharing data between different platforms, which may lead to insufficient data for recommendation systems and further affects recommendation performance.

Federated learning is a privacy-preserving distributed learning scheme proposed by Google\cite{2016Communication}, which enables participants to collaboratively train a machine learning model by sharing intermediate parameters (e.g., model parameters, gradients) instead of their real data. Therefore, combining federated learning with recommendation systems becomes a promising solution for privacy-preserving recommendation systems. In this paper, we term it federated recommendation system (FedRS).

\subsection{Challenges}
While FedRS avoids direct exposure of real user data and provides a privacy-aware paradigm for model training, there are still some core challenges that need to be addressed.

\textbf{Challenge 1: Privacy concerns for users.} Privacy protection is often the major goal of FedRS. In FedRS, each participant jointly trains a global recommendation model by sharing intermediate parameters instead of their real user-item interaction data, which makes an important step towards privacy-preserving recommendation systems. However, a curious server can still infer user ratings and user interaction behaviors from the uploaded intermediate parameters\cite{FedMF}\cite{FedRec}. Besides, FedRS also faces the risk of privacy leakage when integrating auxiliary information (e.g., social features) to improve recommendation performance.

\textbf{Challenge 2: Security attacks on FedRS.} In federated recommendation scenarios, participants may be malicious, and they can poison their local training samples or uploaded intermediate parameters to attack the security of FedRS. They can increase the exposure of specific products for profit\cite{PipAttack}, or destroy the overall recommendation performance of competing companies\cite{FedAttack}. To ensure the fairness and performance of recommendations, FedRS must have the ability to detect and defend against poison attacks from participants.

\textbf{Challenge 3: Heterogeneity in FedRS.} FedRS also faces the problem of system heterogeneity, statistical heterogeneity and privacy heterogeneity during the collaborative training by multiple clients. When training recommendation model locally, due to the difference in storage, computing and communication capabilities, the clients with limited capabilities may become stragglers and further affect the training efficiency. Besides, data (e.g., user attributes, ratings and interaction behavior) in different clients is usually not independent and identically distributed (Non-IID), and training a consistent global recommendation model for all users can't achieve the personalization of recommendation results. Moreover, in realistic applications, users often have different privacy needs and adopt different privacy settings. So simply using the same privacy budgets for users will bring unnecessary loss of recommendation accuracy and efficiency.

\textbf{Challenge 4: Communication costs during FedRS model training and inference.}
To achieve satisfactory recommendation performance, clients need to communicate with the central server for multiple rounds. However, real-world recommendation systems are usually built on complex deep learning models and millions of intermediate parameters need to be communicated\cite{liao2016clustering}. 
In addition, clients must receive a large amount of item data from the server to generate recommendation results locally. Therefore, clients may be hard to afford severe communication costs, which greatly limits the application of FedRS in large-scale recommendation scenarios. 

\subsection{Related Surveys}
There are many surveys that have focused on recommendation systems or federated learning. For examples, Adomavicius $et\ al.$ \cite{2005Toward} provide a detailed categorization of recommendation methods and introduce various limitations of each method. Yang $et\ al.$\cite{FL_Concept_Applications} give the definition of federated learning and discuss its architectures and applications. And Li $et\ al.$\cite{FL_Challenges} summarize the unique characteristics and challenges of federated learning. Besides, there are also some surveys on the privacy and security of federated learning. For examples, Viraaji $et\ al.$\cite{MOTHUKURI2021619} identify and evaluate the privacy threats and security vulnerabilities in federated learning. And Lyu $et\ al.$\cite{2020Privacy} comprehensively explore the assumptions, reasons, principles and differences of the current attacks and defenses in the privacy and robustness fields of federated learning. However, the existing surveys usually treat recommendation systems and federated learning separately, and few work surveyed specific problems in FedRS\cite{FedRecSysSurvey}. Yang $et\ al.$\cite{FedRecSysSurvey} categorize FedRS from the aspect of the federated learning and discuss the algorithm-level and system-level challenges for FedRS. However, they do not provide comprehensive methods to address privacy, security, heterogeneity, and communication costs challenges.

\begin{figure*}[!t]
\centering
\includegraphics[width=1\textwidth]{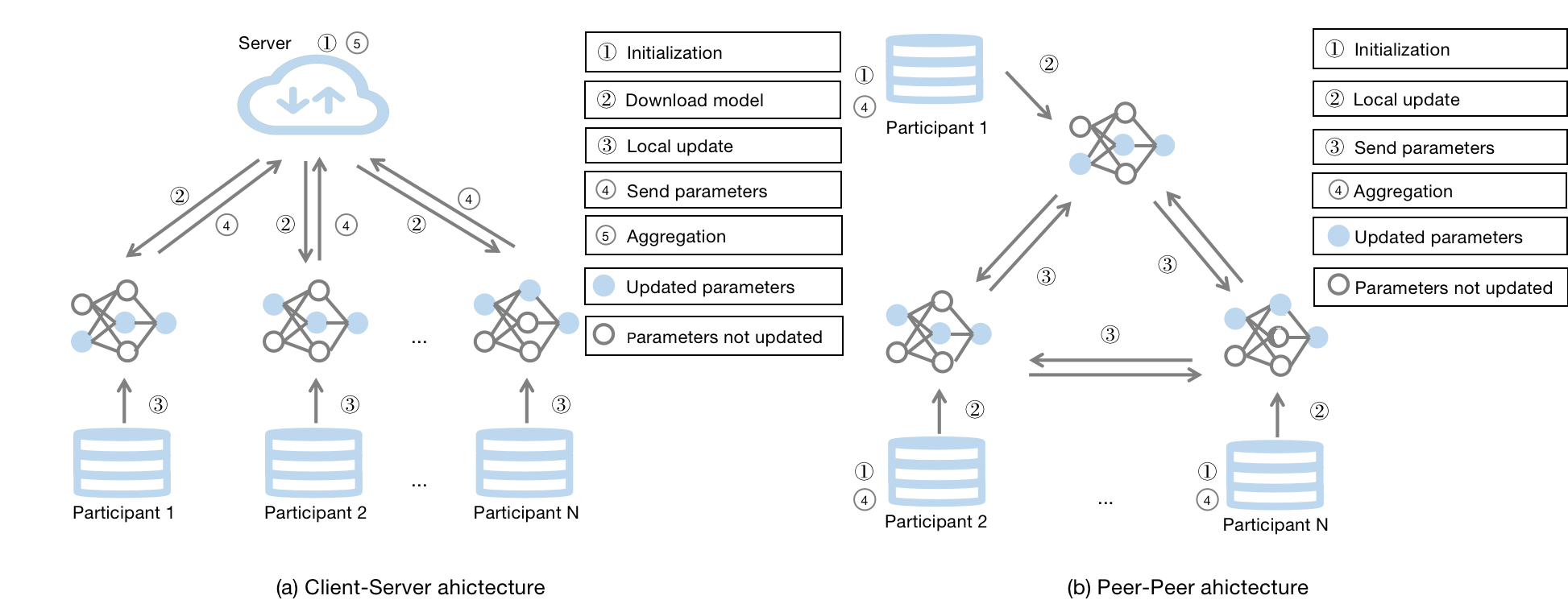}
\caption{Communication architecture of FedRS.}
\label{fig:architecture}
\end{figure*} 
\subsection{Our Contribution}
Compared with the previous surveys, this paper makes the following contributions: Firstly, we provide a comprehensive overview of FedRS from the perspectives of definition, communication architectures and categorization. Secondly, we summarize the state-of-the-art studies of FedRS in terms of privacy, security, heterogeneity and communication costs areas. Thirdly, we introduce some applications and public benchmark datasets for FedRS. Fourthly, we discuss the promising future directions for FedRS. 

The rest of the paper is organized as follows: Section~\ref{sec:overview} discusses the overview of FedRS. Section~\ref{sec:privacy}-Section~\ref{sec:communication} summarize the state-of-the-art studies of FedRS from the aspects of privacy, security, heterogeneity and communication costs. Section~\ref{sec:platform} introduces the applications and public benchmark datasets for FedRS. Section~\ref{sec:future} presents some prospective research directions. Finally, Section~\ref{sec:con} concludes this survey.

\section{Overview of Federated Recommendation Systems}
\label{sec:overview}
\subsection{Definition}
FedRS is a technology that provides recommendation services in a privacy-preserving way. To protect user privacy, the participants in FedRS collaboratively train the recommendation model by exchanging intermediate parameters instead of sharing their own real data. In the ideal case, the performance of recommendation model trained in FedRS should be close to the performance of the recommendation model trained in the data- centralized setting, which can be formalized as:
\begin{equation}\label{eq1}
|V_{FED}-V_{SUM}| < \delta.
\end{equation}
where $V_{FED}$ is the recommendation model performance in FedRS , $V_{SUM}$ is the recommendation model performance in traditional recommendation systems for centralized data storage, and $\delta$ is a small positive number. 

\subsection{Communication Architecture}
In FedRS, the data of participants is stored locally, and the intermediate parameters are communicated between the server and participants. There are two major communication architectures used in the study of FedRS, including client-server architecture and peer-peer architecture.

\textbf{Client-Server Architecture}. Client-server architecture is the most common communication architecture used in FedRS, as shown in Fig. \ref{fig:architecture}(a), which relies on a trusted central server to perform initialization and model aggregation tasks. In each round, the server distributes the current global recommendation model to some selected clients. Then the selected clients use the received model and their own data for local training, and send the updated intermediate parameters (e.g., model parameters, gradients) to the server for global aggregation. The client-server architecture requires a central server to aggregate the intermediate parameters uploaded by the clients. Thus, once the server has a single point of failure, the entire training process will be seriously affected\cite{FedAsync}. In addition, the curious server may infer the clients' privacy information through the intermediate parameters, leaving potential privacy concerns\cite{FedMF}.

\textbf{Peer-Peer Architecture}. Considering the single point of failure problem for client-server architecture in FedRS, Hegeds $et\ al.$\cite{2020Decentralized} design a peer-peer communication architecture with no central server involved in the communication process, which is shown in Fig. \ref{fig:architecture}(b). During each communication round, each participant broadcasts the updated intermediate parameters to some random online neighbors in the peer to peer network, and aggregates received parameters into its own global model. In this architecture, the single point of failure and privacy issues associated with a central server can be avoided. However, the aggregation process occurs on each client, which greatly increases the communication and computation overhead for clients\cite{Efficient-FedRec}.
 
\subsection{Categorization}
\begin{figure}[!t]
\centering
\includegraphics[width=0.5\textwidth]{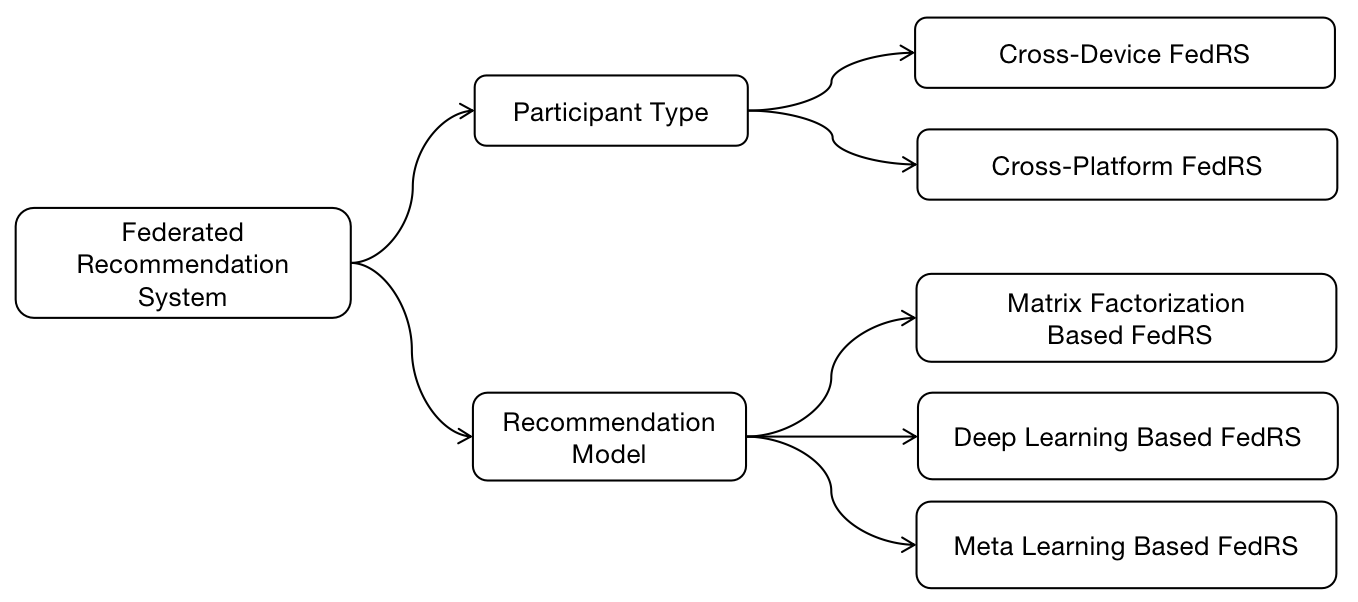}
\caption{Categorization of federated recommendation systems.}
\label{fig:categorization}
\end{figure}

In FedRS, the participants are responsible for the local training process as the data owners. They can be different mobile devices or data platforms. Considering the unique properties of different participant types, FedRS usually have different application scenarios and designs. Besides, there are also some differences between different recommendation models in the federation process. Thus, we summarize the current FedRS and categorize them from the perspectives of participant type and recommendation model. Fig. \ref{fig:categorization} shows the summary of the categorization of FedRS.

\subsubsection{Participant Type}
Based on the type of participants, FedRS can be categorized into cross-device FedRS and cross-platform FedRS.

\textbf{Cross-device FedRS}. In cross-device FedRS, different mobile devices are usually treated as participants \cite{FCF}\cite{FedRec}. The typical application of cross-device FedRS is to build a personal recommendation model for users without collecting their local data. In this way, users can enjoy recommend services while protecting their private information. The number of participants in cross-device FedRS is relatively large and each participant keeps a small amount of data. Considering the limited computation and communication abilities of mobile devices, cross-device FedRS cannot handle very complex training tasks. Besides, due to the power and the network status, mobile devices may drop out of the training process. Thus, the major challenges for cross-device FedRS are how to improve the efficiency and deal with the straggler problem of devices during the training process.

\textbf{Cross-platform FedRS}. In cross-platform FedRS, different data platforms are usually treated as participants who want to collaborate to improve recommendation performance while meeting regulation and privacy constraints\cite{chen2020secure}\cite{WuChuhan2021FedCTR}\cite{2021Horizontal}. For example, to improve the recommendation performance, recommendation systems often integrate data from multiple platforms (e.g., e-commercial platforms, social platforms)\cite{9853237}. However, due to privacy and regulation concerns, the different data platforms are often unable to directly share their data with each other. In this scenario, cross-platform FedRS can be used to collaboratively train recommendation models between different data platforms without directly exchanging their users' data. Compared to cross-device FedRS, the number of participants in cross-platform FedRS is relatively small, and each participant owns relative large amount of data. An important challenge for cross-platform FedRS is how to design a fair incentive mechanism to measure contributions and benefits of different data platforms. Besides, it is hard to find a trusted server to manage training process in cross-platform FedRS, so a peer to peer communication architecture can be a good choice in this case.

\subsubsection{Recommendation Model}
According to the different recommendation models used in FedRS, FedRS can be categorized into matrix factorization based FedRS, deep learning based FedRS and meta learning based FedRS.

\textbf{Matrix factorization based FedRS}. Matrix factorization \cite{koren2009matrix} is the most common model used in FedRS, which formulates the user-item interaction or rating matrix $R\in \mathbb{R}^{N\times M}$ as a linear combination of user profile matrix  $U\in \mathbb{R}^{N\times K}$ and item profile matrix $V\in \mathbb{R}^{M\times K}$: 
\begin{equation}\label{eq2}
R = UV^T.
\end{equation}
then uses the learned model to recommend new items to the user according to the predicted value. In matrix factorization model based FedRS, the user factor vectors are stored and updated locally on the clients, and only the item factor vectors\cite{dolui2019towards} or the gradients of item factor vectors\cite{FCF}\cite{FedRec}\cite{FedMF}\cite{SharedMF} are uploaded to the server for aggregation. Matrix factorization model based FedRS can simply and effectively capture user tastes with the interaction and rating information between users and items. However, it still has many limitations such as sparsity (the number of ratings to be predicted is much smaller than the known ratings) and cold-start (new users and new items lack ratings) problems\cite{2005Toward}.

\textbf{Deep learning based FedRS}. To learn more complex representations of users and items and improve recommendation performance, deep learning technology has been widely used in recommendation systems. However, as privacy regulations get stricter, it becomes more difficult for recommendation systems to collect enough user data to build a high performance deep learning model. To make full use of user data while meeting privacy regulations, many effective deep learning model based FedRS have been proposed\cite{FedNCF}\cite{FedGNN}\cite{imran2022refrs}. Considering different model structures, deep learning model based FedRS usually adopt different model update and intermediate parameter transmit processes. For examples, Perifanis $et\ al.$\cite{FedNCF} propose a federated neural collaborative filtering (FedNCF) framework based on NCF\cite{he2017neural}. In FedNCF, the clients locally update the network weights as well as the user and item profiles, then upload the item profile and network weights after masking to the server for aggregation. Wu $et\ al.$\cite{FedGNN} propose a federated graph neural network (FedGNN) framework based on GNN. In FedGNN, the clients locally train GNN models and update the user/item embeddings from their local sub-graph, then send the perturbed gradients of GNN model and item embedding to the central server for aggregation. Besides, Huang $et\ al.$\cite{FL-MV-DSSM} propose a federated multi-view recommendation framework based on Deep Structured Semantic Model (DSSM\cite{huang2013learning}). In FL-MV-DSSM, each view $i$ locally trains the user and item sub-models based on their own user data and local shared item data, then send the perturbed gradients of both user and item sub-models to server for aggregation. Although deep learning model based FedRS achieve outstanding performance in terms of accuracy, the massive model parameters of deep learning models bring huge computation and communication overhead to the clients, which presents a serious challenge for real industrial recommendation scenarios.

\textbf{Meta learning based FedRS}. The most of existing federated recommendation studies are built on the assumption that data distributed on each client is independent and identically (IID). However, learning a unified federated recommendation model often performs poorly when handling Non-IID and highly personalized data on clients. Meta learning model can quickly adapt to new tasks while maintaining good generalization ability \cite{chen2019closer}, which makes it particularly suitable for FedRS. In meta learning model based FedRS, the server aggregates the intermediate parameters uploaded by clients to learn a model parameter initialization, and the clients fine-tune the initialed model parameters in the local training phase to fit their local data\cite{nichol2018first}\cite{FedMeta}. In this way, meta learning model based FedRS can adapt the clients' local data to provide more personalized recommendations. Although the performance of meta learning model based FedRS are generally better than learning a unified global model, the private information leakage can still occur during the learning process of model parameter initialization \cite{nichol2018first}.

\section{Privacy of Federated recommendation Systems}
\label{sec:privacy}
In the model training process of FedRS, the user data is stored locally and only the intermediate parameters are uploaded to a server, which can further protect user privacy while keeping recommendation performance. Nevertheless, several research works show that the central server can still infer some sensitive information based on intermediate parameters. For examples, a curious server can identify items the user has interacted with according to the non-zero gradients sent by the client\cite{FedGNN}. Besides, the server can also infer the user ratings as long as obtaining the user uploaded gradients in two consecutive rounds\cite{FedMF}. To further protect the privacy of FedRS, many studies have incorporated other privacy protection mechanisms into the FedRS, including pseudo items, homomorphic encryption, secret sharing and differential privacy. This section introduces the application of each privacy mechanism used in FedRS, and compare their advantages and limitations.

\subsection{Pseudo Items}
To prevent the server from inferring the set of items that users have interacted with based on non-zero gradients, some studies utilize pseudo items to protect user interaction behaviors in FedRS. The key idea of pseudo items is that the clients not only upload gradients of items that have been interacted with but also upload gradients of some sampled items that have not been with. 

For example, Lin $et\ at.$\cite{FedRec} propose a federated recommendation framework for explicit feedback scenario named FedRec, in which they design an effective hybrid filling strategy to generate virtual ratings of unrated items by the following equation:
\begin{equation}
r_{ui}^{'}=\left\{
\begin{aligned}
\frac{\sum_{k=1}^{m}y_{uk}r_{uk}}{\sum_{k=1}^{m}y_{uk}}, \  t<T_{predict}\\
\hat{r}_{ui},\  t\geq T_{predict}
\end{aligned}
\right.
\end{equation}
where $t$ denotes the number of current training iteration, and $T_{predict}$ denotes the iteration number when choosing the average value or predict value as virtual rating value to a sampled item $i$. However, the hybrid filling strategy in FedRec introduces extra noise to the recommendation model, which inevitably affects the model performance. To tackle this problem, Feng $et\ at.$\cite{FedRec++} design a lossless version of FedRec named FedRec++. FedRec++ divides clients into ordinary clients and denoising clients. The denoising clients collect noisy gradients from ordinary clients and send the summation of the noisy gradients to the server to eliminate the gradient noise.

Although pseudo items can effectively protect user interaction behaviors in FedRS, it does not modify the gradients of rated items. The curious server can still infer user ratings on the gradients uploaded by users\cite{FedMF}.

\subsection{Homomorphic Encryption}
To further protect the user ratings in FedRS, many studies attempt to encrypt intermediate parameters before uploading them to the server. Homomorphic encryption mechanism allows mathematical operation on encrypted data\cite{2017A}, so it is well suited for the intermediate parameters upload and aggregation processes in FedRS. 

For example, Chai $et\ at.$\cite{FedMF} propose a secure federated matrix factorization framework named FedMF, in which clients use Paillier homomorphic encryption mechanism \cite{1999Public} to encrypt the gradients of item embedding matrix before uploading them to the server, and the server aggregates gradients on the cipher-text. Due to the characteristics of homomorphic encryption, FedMF can achieve the same recommendation accuracy as traditional matrix factorization. However, FedMF causes serious computation overheads since all computation operations are performed on the ciphertext and most of system’s time is spent on server updates. Besides, FedMF assumes that all participants are honest and will not leak the secret key to the server, which is hard to guarantee in reality. Moreover, Zhang $et\ at.$\cite{zhang2021vertical} propose a federated recommendation method (CLFM-VFL) for vertical federated learning scenarios where participants have more overlapping users but fewer overlapping features of users. CLFM-VFL uses homomorphic encryption to protect the gradients of user hidden vectors for each participant and cluster the users to improve recommendation accuracy and reduce matrix dimension.

Besides, many studies also utilize homomorphic encryption mechanism to integrate private information from other participants to improve recommendation accuracy \cite{FedGNN}\cite{FedPOIRec}. For examples, Wu $et\ al.$\cite{FedGNN} use homomorphic encryption mechanism to find the anonymous neighbors of users to expand the local user-item graph. Perifanis $et\ al.$\cite{FedPOIRec} use Cheon-Kim-Kim-Song (CKKS) fully homomorphic encryption mechanism\cite{2017Homomorphic} to incorporate learned parameters between user's friends after the global model is generated.

Homomorphic encryption mechanism based FedRS can effectively protect user ratings while maintaining recommendation accuracy. Besides, it can prevent privacy leaks when integrating information from other participants. However, homomorphic encryption brings huge computation costs during operation process. And it is also a serious challenge to keep the secret key not be obtained by the server or other malicious participants.

\begin{table*}[!htb]
\centering
\caption{Comparison between different privacy mechanism.} 
\renewcommand{\arraystretch}{2}
\label{table:privacy}
\setlength{\tabcolsep}{3mm}{ 
\begin{tabular}{|l|l|l|c|l|}
\hline
Privacy Mechanisms & Ref & Main Protect Object & Accuracy Loss & Communication/Computation Costs\\
\hline

Pseudo Items & \cite{FedRec}\cite{FedRec++}\cite{FedGNN}\cite{FeSoG}\cite{FR-FMSS}& Interaction Behaviors&\Checkmark& Low Costs\\ \hline

\multirow{3}*{Homomorphic Encryption} & \cite{FedMF} & Ratings&\multirow{3}*{\XSolidBrush}&\multirow{3}*{High Computation Costs}\\

& \cite{FedGNN} & High-order Graph & & \\ 
& \cite{FedPOIRec} & Social Features & & \\\hline

Secret Sharing &\cite{SharedMF}\cite{FR-FMSS}& Ratings&\XSolidBrush& High Communication Costs\\ \hline

Local Differential Privacy &\cite{dolui2019towards}\cite{FedGNN}\cite{FeSoG}& Ratings&\Checkmark& Low Costs\\

\hline
\end{tabular}}
\end{table*}

\subsection{Secret Sharing}
As another encryption mechanism used in FedRS, secret sharing mechanism breaks intermediate parameters up into multiple pieces, and distributes the pieces among participants, so that only when all pieces are collected can reconstruct the intermediate parameters.

For example, Ying \cite{SharedMF} proposes a secret sharing based federated matrix factorization framework named ShareMF. The participants divide the item matrix gradients $g^{plain}$ into several random numbers that meet:
\begin{equation}\label{eq6}
g^{plain}=g^{sub1}+g^{sub2}+...+g^{subt}.
\end{equation}
Each participant keeps one of the random numbers and sends the rest to $t-1$ sampled participants, then uploads the sum of received and kept numbers as hybrid gradients to the server for aggregation. ShareMF protects the user ratings and interaction behaviors from being inferred by the server, but the rated items can still be leaked to other participants who received the split numbers. To tackle this problem, Lin $et\ al.$ \cite{FR-FMSS} combine secret sharing and pseudo items mechanisms to provide a stronger privacy guarantee.

Secret sharing mechanism based FedRS can protect user ratings while maintaining recommendation accuracy, and have lower computation costs compared to homomorphic encryption based FedRS. But the exchange process of pieces between participants greatly increases the communication costs.

\subsection{Local Differential Privacy}
Considering the huge computation or communication costs caused by encryption based mechanisms, many studies try to use perturbation based mechanisms to adapt to large-scale FedRS for industrial scenarios. Local differential privacy (LDP) mechanism allows statistical computations while guaranteeing each individual participant’s privacy\cite{2012Calibrating}, which can be used to perturb the intermediate parameters in FedRS. 

For example, Dolui $et\ al.$\cite{dolui2019towards} propose a federated matrix factorization framework, which applies differential privacy on item embedding matrix before sending it to the server for weighted average. However, the server can still infer which items the user has rated just by comparing the changes in item embedding matrix.

In order to achieve more comprehensive privacy protection during model training process, Wu $et\ al.$ \cite{FedGNN} combine pseudo items and LDP mechanisms to protect both user interaction behaviors and ratings in FedGNN. Firstly, to protect user interaction behaviors in FedGNN, the clients randomly sample $N$ items that they have not interacted with, then generate the virtual gradients of item embeddings by using the same Gaussian distribution as the real embedding gradients. Secondly, to protect user ratings in FedGNN, the clients apply a LDP module to clip the gradients according to their L2-norm with a threshold $\delta$ and perturb the gradients by adding zero-mean Laplacian noise. The LDP module of FedGNN can be formulated as follow:
\begin{equation}\label{eq5}
g_i=clip(g_i,\delta)+Laplace(0,\lambda).
\end{equation}
where $\lambda$ is the Laplacian noise strength. However, the gradient magnitude of different parameters varies during training process, thus it is usually not appropriate to perturb gradients at different magnitudes with a constant noise strength. So Liu $et\ al.$ \cite{FeSoG} propose to add dynamic noise according to the gradients, which can be formulated as follow:
\begin{equation}\label{eq6}
g_i=clip(g_i,\delta)+Laplace(0,\lambda\cdot mean(g_i)).
\end{equation}

Local differential privacy mechanism doesn't bring heavy computation and communication overhead to FedRS, but the additional noise inevitably affects the performance of the recommendation model. Thus, in the actual application scenario, we must consider the trade-off between privacy and recommendation accuracy.

\subsection{Comparison}
To provide a stronger privacy guarantee, many privacy mechanisms (i.e., pseudo items, homomorphic encryption, differential privacy and secret sharing) have been widely used in FedRS, and the comparison between these mechanisms is shown in Table \ref{table:privacy}. Firstly, the main protect objects of these mechanisms are different: pseudo items mechanism is to protect user interaction behaviors, and the rest mechanisms are to protect user ratings. Besides, homomorphic encryption can also integrate data from other participants in a privacy-preserving way. Secondly, homomorphic encryption and secret sharing are both encryption-based mechanisms, and they can protect privacy while keeping accuracy. However, the high computation cost of homomorphic encryption limits it's application in large-scale industrial scenarios. Although the secret sharing mechanism reduces the computation costs, the communication costs increase greatly. Pseudo items and differential privacy mechanisms protect privacy by adding random noise, which has low computation costs and don't bring additional communication costs. But the addition of random noise will inevitably affect model performance to a certain extent.

\section{Security of Federated recommendation Systems}
\label{sec:security}
Apart from privacy leakage problems, traditional recommendation systems for centralized data storage are also vulnerable to poisoning attacks (shilling attacks)\cite{2018Poisoning}\cite{2021Data}\cite{2020Practical}. Attackers can poison recommendation systems and make recommendations as their desire by injecting well-crafted data into the training dataset. But most of these poisoning attacks assume that the attackers have full prior knowledge of entire training datasets. Such an assumption may be not valid for FedRS since the data in FedRS is distributed and stored locally for each participant. Thus, FedRS provides a stronger security guarantee than traditional recommendation systems. However, the latest studies indicate that attackers can still conduct poisoning attacks on FedRS with limited prior knowledge\cite{PipAttack}\cite{FedAttack}\cite{2022FedRecAttack}\cite{rong2022poisoning}. In this section, we summarize some novel poisoning attacks against FedRS and provide some defense methods.

\begin{table*}[!htb]
\centering
\caption{Representative works on the security of FedRS. RA refers to robust aggregation and AD refers to anomaly detection.} 
\renewcommand{\arraystretch}{2}
\label{table:security}
\setlength{\tabcolsep}{3mm}{
\begin{tabular}{|ll|c|c|c|c|c|c|l|}
\hline
\multirow{2}*{Works} & \multirow{2}*{Ref} & \multicolumn{2}{c}{Attack Type} & \multicolumn{2}{|c}{Poison Object} & \multicolumn{2}{|c|}{Defense Type}&\multirow{2}*{Goal}\\
\cline{3-8}
&& Target & Untarget & Model & Data & RA & AD &

\\ \hline
PipAttack&\cite{PipAttack}&\Checkmark &&\Checkmark& &&&Increase/decrease popularity of target items.\\ 

FedRecAttack&\cite{2022FedRecAttack}&\Checkmark &&\Checkmark& &&&Increase/decrease popularity of target items. \\

A-ra/A-hum&\cite{rong2022poisoning}&\Checkmark &&\Checkmark& &&&Increase/decrease popularity of target items. \\ 

FedAttack&\cite{FedAttack}&&\Checkmark&&\Checkmark&&& Degrade the overall performance of FedRS. \\ \hline

Median&\cite{Median}&&&&&\Checkmark&& Guarantee global model convergence. \\ 

Trimmed-Mean&\cite{Median}&&&&&\Checkmark&& Guarantee global model convergence. \\ 

(Multi-)Krum&\cite{Krum}&&&&&\Checkmark&& Guarantee global model convergence. \\ 

Bulyan&\cite{2018The}&&&&&\Checkmark&& Guarantee global model convergence. \\ 

Norm-Bounding&\cite{2019Can}&&&&&\Checkmark&& Guarantee global model convergence. \\ 

A-FRS&\cite{A-FRS}&&&&&\Checkmark&& Guarantee global model convergence. \\ 
FSAD&\cite{FSAD}&&&&&&\Checkmark& Identify and filter poisoned parameters. \\

\hline
\end{tabular}}
\end{table*}

\subsection{Poisoning Attacks}
According to the goal of attacks, the poisoning attacks against FedRS can be categorized into targeted attacks and untargeted attacks as shown in Table \ref{table:security}. 

\subsubsection{Target Poisoning Attacks}
The goal of target attacks on FedRS is to increase or decrease the exposure chance of specific items, which are usually driven by financial profit. For example, Zhang $et\ al.$\cite{PipAttack} propose a poisoning attack for item promotion (PipAttack) against FedRS by utilizing popularity bias. To boost the rank score of target items, PipAttack uses popularity bias to align target items with popular items in the embedding space. Besides, to avoid damaging recommendation accuracy and being detected, PipAttack designs a distance constraint to keep modified gradients uploaded by malicious clients close to normal ones. 

In order to further reduce the degradation of recommendation accuracy caused by targeted poisoning attacks, and the proportion of malicious clients needed to ensure the attack effectiveness, Rong\cite{2022FedRecAttack} propose a model poisoning attack against FedRS (FedRecAttack), which makes use of a small proportion of public interactions to approximate the user feature matrix, then uses it to generate poisoned gradients.

Both PipAttack and FedRecAttack rely on some prior knowledge. For example, PipAttack assumes the attack can access popularity information, and FedRecAttack assumes the attacker can get public interactions. So the attack effectiveness is greatly reduced in the absence of prior knowledge, which makes both attacks not generic in all FedRS. To make attackers conduct effective poisoning attacks to FedRS without prior knowledge, Rong $et\ al.$\cite{rong2022poisoning} design two methods (i.e., random approximation and hard user mining) for malicious clients to generate poisoned gradients. In particular, random approximation (A-ra) uses Gaussian distribution to approximate normal users’ embedding vectors, and hard user mining (A-hum) uses gradient descent to optimize users’ embedding vectors obtained by A-ra to mine hard users. In this way, A-hum can still effectively attack FedRS with extremely small proportion of malicious users.

\subsubsection{Untarget Poisoning Attacks}
The goal of untarget attacks on FedRS is to degrade the overall performance of the recommendation model, which are usually conducted by competing companies. For example, Wu $et\ al.$ \cite{FedAttack} propose an untargeted poisoning attack to FedRS named FedAttack, which uses a globally hard sampling technique\cite{2020Hard} to subvert model training process. More specifically, after inferring the user's interest from local user profiles, the malicious clients select candidate items that best match the user's interest as negative samples, and select candidate items that least match the user's interest as positive samples. FedAttack only modifies training samples, and the malicious clients are also similar to normal clients with different interests, thus FedAttack can effectively damage the performance of FedRS even under defense.

\subsection{Defense Methods}
To reduce the influence of poisoning attacks on FedRS, many defense methods have been proposed in the literature, which can be classified into robust aggregation and anomaly detection.

\subsubsection{Robust Aggregation}
The goal of robust aggregation is to guarantee global model convergence when up to 50\% of participants are malicious\cite{2020Privacy}, which selects statistically more robust values rather than the mean values of uploaded intermediate parameters for aggregation.

\textbf{Median\cite{Median}.} Median selects the median value of each updated model parameter independently as aggregated global model parameter, which can represent the center of the distribution better. Specifically, the server ranks each $i-th$ parameter of $n$ local model update, and uses the median value as $i-th$ parameter of the global model.

\textbf{Trimmed-Mean\cite{Median}.} Trimmed-Mean removes the maximum and minimum values of each updated model parameter independently, and then takes the mean value as aggregated global model parameter. Specifically, the server ranks each $i-th$ parameter of $n$ local model update, removes $\beta$ smallest and $\beta$ largest values, and uses the mean value of remaining $n-2\beta$ as $i-th$ parameter of global model. In this way, Trimmed-Mean can effectively reduce the impact of outliers.

\textbf{Krum and Multi-Krum\cite{Krum}.} Krum selects a local model that is the closest to the others as the global model. Multi-Krum selects multiple local models by using Krum, then aggregates them into a global model. In this way, even if the selected parameter vectors are uploaded by malicious clients, their impact is still limited because they are similar to other local parameters uploaded by normal clients.

\textbf{Bulyan\cite{2018The}.} Bulyan is a combination of Krum and Trimmed-Mean, which iteratively selects $m$ local model parameter vectors through Krum, and then performs Trimmed-Mean on these $m$ parameter vectors for aggregation. With high dimensional and highly non-convex loss functions, Bulyan can still converge to effectual models.

\textbf{Norm-Bounding\cite{2019Can}.} Norm-Bounding clips the received local parameters to a fixed threshold, then aggregates them to update the global model. Norm-Bounding can limit the contribution of each local model updates so as to mitigate the affect of poisoned parameters on the aggregated model.

\textbf{A-FRS\cite{A-FRS}.} A-FRS utilizes gradient-based Krum instead of model parameter-based Krum to filter malicious clients in momentum-based FedRS. A-FRS theoretically guarantees that if the selected gradient is close to the normal gradient, the momentum and model parameters will also be close to the normal momentum and model parameters.

Although these robust aggregation strategies provide convergence guarantees to some extent, most of them (i.e., Bulyan, Krum, Median and Trimmed-mean) greatly degrade the performance of FedRS. Besides, some noval attacks(i.e., PipAttack, FedAttack)\cite{PipAttack}\cite{FedAttack} utilize well-designed constraints to approximate the patterns of normal users and circumvent defenses, which further increases the difficulty of defense.

\subsubsection{Anomaly Detection}
The purpose of anomaly detection strategy is to identify the poisoned model parameters uploaded by malicious clients and filter them during the global model aggregation process. For example, Jiang $et\ al.$\cite{FSAD} propose an anomaly detection strategy named federated shilling attack detector (FSAD) to detect poisoned gradients in federated collaborative filtering scenarios. FSAD extracts 4 novel features according to the gradients uploaded by clients, then uses the gradient-based features to train a semi-supervised bayes classifier so as to identify and filter the poisoned gradients. However, in FedRS, the interests of different users vary widely, thus the parameters they uploaded are usually quite different, which increases the difficulty of anomaly detection \cite{2022FedRecAttack}.

\begin{figure*}[!t]
\centering
\includegraphics[width=1\textwidth]{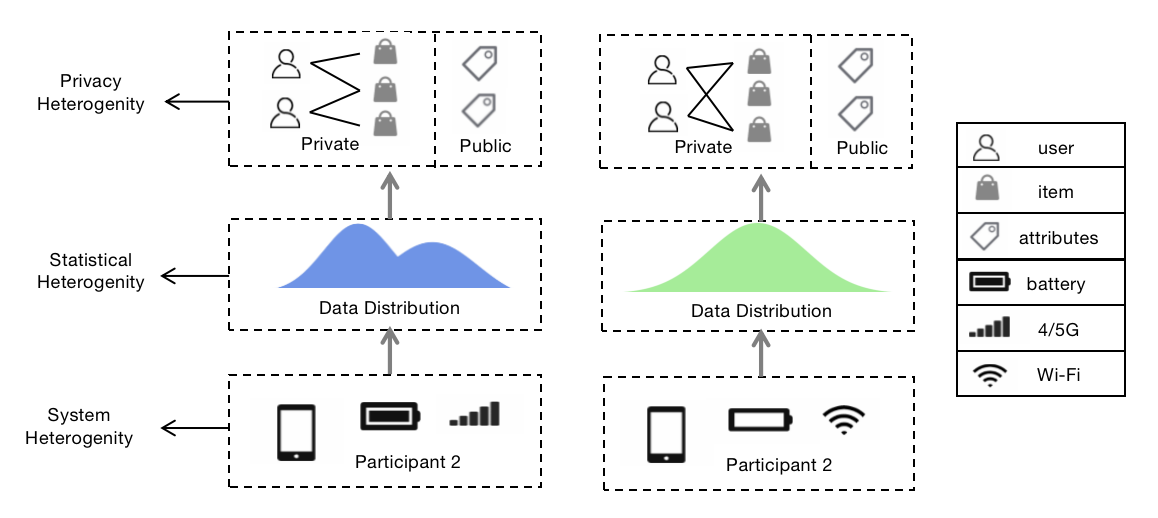}
\caption{Heterogeneity of federated recommendation systems.}
\label{fig:heterogeneity}
\end{figure*}

\section{Heterogeneity of Federated Recommendation Systems}
\label{sec:heterogeneity}
Compared with traditional recommendation systems, FedRS face more severe challenges in terms of heterogeneity, which are mainly reflected in system heterogeneity, statistical heterogeneity and model heterogeneity, as shown in Fig. \ref{fig:heterogeneity}. System heterogeneity refers to client devices have significantly different storage, computation, and communication capabilities. Devices with limited capabilities greatly affect training efficiency, and further reduce the accuracy of the global recommendation model.\cite{kairouz2021advances}; Statistical heterogeneity refers to the data collected by different clients is usually not independent and identically distributed (non-IID). As a result, simply training a single global model is difficult to generalize to all clients, which affects the personalization of recommendations\cite{0Personalized}; Privacy heterogeneity means that the privacy constraints of different users and information vary greatly, so simply treating them with the same privacy budgets will carry unnecessary costs\cite{HPFL}. This section introduces some effective approaches to address the heterogeneity of FedRS.

\subsection{System Heterogeneity}
In FedRS, the hardware configuration, network bandwidth and battery capacity of participating clients vary greatly, which results in diverse computing capability, communication speed, and storage capability\cite{FL_Challenges}. During the training process, the clients with limited capacity could become stragglers, and even drop out of current training due to network failure, low battery and other problems\cite{FedAsync}. The system heterogeneity significantly delays the training process of FedRS, further reducing the recommendation accuracy of the global model. To make the training process compatible with different hardware structures and tolerate the straggling and exit issues of clients, the most common methods are asynchronous communication\cite{FedSA}\cite{FedAsync} and clients selection\cite{FedCS}.

\textbf{Asynchronous communication.} Considering the synchronous communication based federated learning must wait for straggler devices during the aggregation process, many asynchronous communication strategies are presented to improve training efficiency. For examples, FedSA\cite{FedSA} proposes a semi-asynchronous communication method, where the server aggregates the local models based on their arrival order of each round. FedAsync\cite{FedAsync} uses a weighted average strategy to aggregate the local models based on staleness, which assigns less weight to delayed feedback in the update process. 

\textbf{Clients selection}. Client selection approach selects clients for updates based on resource constraints so that the server can aggregate as many local updates as possible at the same time. For example, in FedCS\cite{FedCS}, the server sends a resource request to each client so as to get their resource information, then estimates the required time of model distribution, updating and uploading processes based on the resource information. According to the estimated time, the server determines which clients can participant in the training process.

\subsection{Statistical Heterogeneity} 
Most of the existing federated recommendation studies are built on the assumption that data in each participant is independent and identically distributed (IID). However, the data distribution of each client usually varies greatly, hence training a consistent global model is difficult to generalize to all clients under non-IID data and inevitably neglects the personalization of clients\cite{HPFL}. To address the statistical heterogeneity problem of FedRS, many effective strategies have been proposed, which are mainly based on meta learning\cite{FedMeta}\cite{2021FastAdapt} and clustering\cite{jie2022personalized}\cite{luo2022personalized}.

\textbf{Meta learning.} As known as “learning to learn”, meta learning technology aims to quickly adapt the global model learned by other tasks to a new task by using only a few samples\cite{chen2019closer}. The rapid adaptation and good generalization abilities make it particularly well-suited for building personalized federated recommendation models. For examples, FedMeta\cite{FedMeta} uses Model-Agnostic Meta-Learning (MAML)\cite{2017Model} algorithm to learn a well-initialized model that can be quickly adapted to clients, and effectively improve the personalization and convergence of FedRS. However, FedMeta needs to compute the second-order gradients, which greatly increases computation costs. Besides, the data split process also brings a huger challenge for clients with limited samples. Based on FedMeta, Wang $et\ al.$\cite{2021FastAdapt} propose a new meta learning algorithm called Reptile which applies the approximate first-order derivatives for the meta-learning updates, which greatly reduces the computation overloads of clients. Moreover, Reptile doesn’t need a data split process, which makes it also suitable for clients with limited samples.

\textbf{Clustering.} The core idea of clustering is training personalized models jointly with the same group of homogeneous clients. For examples, Jie $et\ al.$\cite{jie2022personalized} use historical parameter clustering technology to realize personalized federated recommendation, in which the server aggregates local parameters to generate global model parameters and clusters the local parameters to generate clustering parameters for different client groups. Then the clients combine the clustering parameters with the global parameters to learn personalized models. Luo $et\ al.$\cite{luo2022personalized} propose a personalized federated recommendation framework named PerFedRec, which constructs a collaborative graph and integrates attribute information so as to jointly learn the user representations by federated GNN. Based on the learned user representations, clients are clustered into different groups. And each cluster learns a cluster-level recommendation model. At last, each client can obtain a personalized model by merging the global recommendation model, the cluster-level recommendation model, and the fine-tuned local recommendation model. Although clustering based approaches can alleviate statistical heterogeneity, the clustering and combination process greatly increase the computation costs.

\subsection{Privacy Heterogeneity}
In reality, the privacy restrictions of different participants and information vary greatly, thereby using the same high level of privacy budget for all participants and information is unnecessary, which even increases the computation/communication costs and degrades the model performance. 

\textbf{Heterogeneous user privacy}. In order to adapt to the privacy needs of different users, Anelli $et\ al.$\cite{2021FedeRank} present a user controlled federated recommendation framework named FedeRank. FedeRank introduces a probability factor $\pi\in[0,1]$ to control the proportion of interacted item updates and masks the remain interacted item update by setting them to zero. In this way, FedeRank allows users to decide the proportion of data they want to share by themselves, which addresses the heterogeneity of user privacy.

\textbf{Heterogeneous information privacy}. In order to adapt to the privacy needs of different information components, HPFL\cite{HPFL} designs a differentiated component aggregation strategy. To obtain the global public information components, the server directly weighted aggregates the local public components with the same properties. And to obtain the global privacy information components, the user and item representations are kept locally, and the server only aggregates the local drafts without the need to align the presentations. With the differentiated component aggregation strategy, HPFL can safely aggregate components with heterogeneous privacy constraints in user modeling scenarios.

\section{Communication Costs of Federated recommendation Systems}
\label{sec:communication}
To achieve satisfactory recommendation performance, FedRS requires multiple communications between the server and clients. However, the real-world recommendation systems are usually conducted by complexity deep learning models with large model size\cite{2020Understanding}, and millions of parameters need to be updated and communicated \cite{liao2016clustering}, which brings severe communication overload to resource limited clients and further affects the application of FedRS in large-scale industrial scenarios. This section summarizes some optimization methods to reduce communication costs of FedRS, which can be classified into importance-based updating\cite{PPRSF}\cite{Efficient-FedRec}\cite{FCF-BTS}\cite{ai2022fourier}, model compression\cite{2016Federated}\cite{JointRec}, active sampling\cite{FedFast} and one shot learning\cite{2022FedSPLIT}.

\subsection{Importance-based Model Updating}
Importance-based model updating selects important parts of the global model instead of the whole model to update and communicate, which can effectively reduce the communicated parameter size in each round.

For examples, Qin $et\ al.$\cite{PPRSF} propose a federated framework named PPRSF, which uses 4-layers hierarchical structure for reducing communication costs, including the recall layer, ranking layer, re-ranking layer and service layer. In the recall layer, the server roughly sorts the large inventory by using public user data, and recalls a relatively small number of items for each client. In this way, the clients only need to update and communicate the candidate item embeddings, which greatly reduces the communication costs between the server and clients, and the computation costs in the local model training and inference phases. However, the recall layer of PPRSF needs to get some public information about users, which raises certain difficulty and privacy concerns.

Yi $et\ al.$\cite{Efficient-FedRec} propose an efficient federated news recommendation framework called Efficient-FedRec, which breaks the news recommendation model into a small user model and a big news model. Each client only requests the user model and a few news representations involved in their local click history for local training, which greatly reduces the communication and computation overhead. To further protect specific user click history against the server, they transmit the union news representations set involved in a group of user click history by using a secure aggregation protocol\cite{bonawitz2017practical}.

Besides, Khan $et\ al.$\cite{FCF-BTS} propose a multi-arm bandit method (FCF-BTS) to select part of the global model that contains a smaller payload for all clients. The rewards of the selection process are guided by Bayesian Thompson Sampling (BTS)\cite{thompson1933likelihood} approach with Gaussian priors. Experiments show that FCF-BTS can reduce 90\% model payload for highly sparse datasets. Besides, the selection process occurs on the server side, thus avoiding additional computation costs for the clients. But FCF-BTS causes 4\% - 8\% loss in recommendation accuracy.

To achieve a better balance between recommendation accuracy and efficiency, Ai $et\ al.$\cite{ai2022fourier} propose an all-MLP network that uses a Fourier sub-layer to replace the self-attention sub-layer in a Transformer encoder so as to filter noise data components unrelated to the user’s real interests, and adapts an adaptive model pruning technique to discard the noise model components that doesn't contribute to model performance. Experiments show that all-MLP network can significantly reduce communication and computation costs, and accelerates the model convergence.

Importance-based model updating strategies can greatly reduce communication and computation costs at the same time, but only selecting the important parts for updating inevitably reduces the recommendation performance.

\subsection{Model Compression}
Model Compression is a well-known technology in distributed learning\cite{2018ATOMO}, which compresses the communicated parameters per round to be more compact.

For examples, Konen $et\ al.$\cite{2016Federated} propose two methods (i.e., structured updates and sketched updates) to decrease the uplink communication costs under federated learning settings. Structured updates method directly learns updates from a pre-specified structure parameterized using fewer variables. Sketched updates method compresses the full local update using a lossy compression way before sending it to the server. These two strategies can reduce communication costs by 2 orders of magnitude.

To reduce the uplink communication costs in deep learning based FedRS, JointRec \cite{JointRec} combines low-rank matrix factorization\cite{2014Compressing} and 8-bit probabilistic quantization\cite{2017DeepG} methods to compress weight update. Supposing the weight update matrix of client n is $H_n^{a\times b}$, $a\leq b$, low-rank matrix factorization decomposes $H_n^{a\times b}$ into two matrices: $H^{a\times b}_n=U^{a\times k}_n V^{k\times b}_n$, where $k=b/N$ and N is a positive number that influences the compression performance. And 8-bit probabilistic quantization method transforms the position of matrix value into 8-bit value before sending it to the server. Experiments demonstrate that JointRec can realize $12.83 \times$ larger compression ratio while maintaining recommendation performance.

Model compression methods achieve significant results in reducing uplink communication costs. However, the reduction of communication costs sacrifices the computation resources of the clients, so it's necessary to consider the trade-off between computation and communication costs when using model compression.

\subsection{Client Sampling}
In traditional federated learning frameworks\cite{2016Communication}, the server randomly selects clients to participate in the training process and simply aggregates the local models by average, which requires a large number of communications to realize satisfactory accuracy. Client sampling utilizes efficient sampling strategies so as to improve training efficiency and reduce the communication rounds.

For example, Muhammad $et\ al.$ \cite{FedFast} propose an effective sampling strategy named FedFast to speed up the training efficiency of federated recommendation models while keeping more accuracy. FadFast consists of two efficient components: ActvSAMP and ActvAGG. ActvSAMP uses K-means algorithm to cluster users based on their profiles, and samples clients in equal proportions from each cluster. And ActvAGG propagates local updates to the other clients in the same cluster. In this way, the learning process for these similar users is greatly accelerated and the overall efficiency of the FedRS is consequently improved. Experiments show that FedFast reduces communication rounds by 94\% compared to FedAvg\cite{2016Communication}. However, FedFast is faced with the cold start problem because it requires a number of users and items for training. Besides, FedFast needs to retrain the model to support new users and items.

\subsection{One Shot Federated Learning}
The goal of one shot federated learning mechanism is to reduce communication rounds of FedRS\cite{2019One}\cite{kasturi2020fusion}, which limits communication to a single round to aggregate knowledge of local models. For example, Eren $et\ al.$ \cite{2022FedSPLIT} implement an one-shot federated learning framework for cross-platform FedRS named FedSPLIT.  FedSPLIT aggregates model through knowledge distillation\cite{li2020practical}, which can generate client specific recommendation results with just a single pair of communication rounds between the server and clients after a small initial communication. Experiments show that FedSPLIT realizes similar root-mean-square error (RMSE) compared with multi-round communication scenarios, but it is not applicable to the scenario where the participants are individual users.

\section{Applications And Public Benchmark Datasets }\label{sec:platform}
This section introduces the typical applications and public benchmark datasets for FedRS.
\subsection{Applications}
\textbf{Online services.}
Currently, online services have been involved in various fields of our life such as news, movie and music. A large amount of private information of users is collected and stored centrally by service providers, which faces a serious risk of privacy leakage. User data may be sold to third parties by service providers or stolen by external hackers. FedRS can help users enjoy personalized recommendation services while keeping personal privacy, make the service providers more trusted by users, and ensure the recommendation service complies with the regulations. For example, Tan $et\ al.$ \cite{10.1145/3383313.3411528} design a federated recommendation system that implements various popular recommendation algorithms to support lots of online recommendation services, and deploys it on a real-world content recommendation application.

\textbf{Healthcare.}
Healthcare recommendation enables patients to enjoy medical service from mobile applications instead of going to the hospital in person when obtaining satisfactory recommendations. Medical data is quite private and sensitive, which means it is hard to fuse user information from different hospitals or other organizations to improve recommendation quality. In this scenario, FedRS can break down the data silos and utilize these data without compromising patients' privacy. For example, Song $et\ al.$\cite{9657870} develop a telecommunication-joint federated healthcare recommendation platform based on Federated AI Technology Enabler (FATE), which helps healthcare providers improve recommendation performance by complementing common user data (e.g., demographic information, user behaviors and geographic information) from mobile network operators. Besides, this platform designs a federated gradient boosting decision tree (FGDBT) model, improving 9.71\% of precision and 4\% of F1 score for healthcare recommendation. The platform has been deployed on both organizations and applied to online operation.

\textbf{Advertisement.} Advertisement is another significant application of FedRS. Platforms that display advertisements often face the problem of insufficient user data and low click-through rates (CTR) for advertisements. FedRS is able to exploit user data across different platforms in a privacy-preserving way, which can better infer user interest and push advertisements more accurately. For example, Wu $et\ al.$ \cite{WuChuhan2021FedCTR} propose a native advertisement CTR prediction method named FedCTR, which can integrate multi-platform user behaviors (e.g., advertisements click behavior, search behavior and browsing behavior) for user interest modeling with no need for centralized storage.

\textbf{E-commerce.}
Currently, recommendation system plays a significant role in e-commerce platforms  (e.g., Alibaba, Amazon). To provide users with more precise recommendation services, such systems try to integrate more auxiliary information (e.g., user purchasing power, social information). However, these data are usually distributed on different platforms and difficult to access directly due to regulations and privacy concerns. FedRS can address this problem effectively while meeting regulations and privacy.

\textbf{Point-of-Interest.}
Point-of-Interest (POI) recommendation exploits the user's historical check-in data and other modal information (e.g., POI attributes and social information) to recommend suitable POI sets for the user. However, the user's check-in data is very sensitive and sparse, and users are often reluctant to share their context information due to privacy concerns. FedRS can effectively address the data sparsity problem in a privacy-preserving way, which is quite beneficial for POI recommendation\cite{FL&PP-POI}.

\subsection{Public Benchmark Datasets}

\textbf{MovieLens\cite{10.1145/2827872}.}
MovieLens rating datasets were published by GroupLens, which consist of user, movie, rating and timestamp information. MovieLens-100K contains 100,000 ratings from 943 users for 1682 movies, and MovieLens-1M contains 1,000,209 ratings from 6,040 users for 3,952 movies.

\textbf{FilmTrust\cite{guo2013novel}.} 
FilmTrust is a movie rating dataset crawled from the FilmTrust website. The dataset contains 35,497 ratings from 1,508 users for 2,071 films.

\textbf{Foursquare\cite{10.1145/2814575}.} 
Foursquare dataset is a famous benchmark dataset to evaluate POI recommendation models collected from Foursquare. The dataset contains 22,809,624 global-scale check-ins by 114,324 users on 3,820,891 POIs with 363,704 social relationships.

\textbf{Epinions\cite{10.1145/2124295.2124309}.} 
Epinions dataset is an online social network built from a consumer review site Epinions.com, which consists of user ratings and trust social network information. The dataset contains 188,478 ratings from 116,260 users for 41,269 items.

\textbf{Mind\cite{wu2020mind}.}
Mind is a large-scale dataset for news recommendation collected from anonymous behavior logs of Microsoft News website, which contains about 160,000 English news articles and more than 15 million impression logs generated by 1 million users.

\textbf{LastFM\cite{10.1145/2039320}.} 
LastFM dataset was collected from Last.fm online music system, which consists of tagging, music artist listening, and social relationship information. The dataset contains 92,834 listening counts of 17,632 music artists by 1,892 users.

\textbf{Book-Crossing\cite{ziegler2005improving}.} 
Book-Crossing dataset is a 4-week crawl dataset from the Book-Crossing community. It contains 1,149,780 ratings (explicit / implicit) for 271,379 books by 278,858 anonymous users with demographic information.

\section{Future Directions}\label{sec:future}
This section presents and discusses many prospective research directions in the future. Although some directions have been covered in the above sections, we believe they are necessary for FedRS, and need to be further researched.

\textbf{Decentralized FedRS.} Most current FedRS are based on client-server communication architecture, which faces single-point-of-failure and privacy issues caused by the central server\cite{lyu2020towards}. While much work has been devoted to decentralized federated learning \cite{lyu2020democratise} \cite{9826687}, few decentralized FedRS have been studied. A feasible solution is to replace client-server communication architecture with peer-peer communication architecture to achieve fully decentralized federated recommendation. For example, Hegeds $et\ al.$\cite{2020Decentralized} propose a fully decentralized matrix factorization framework based on gossip learning \cite{ormandi2013gossip}, where each participant sends their copy of the global recommendation model to random online neighbors in the peer to peer network. In addition, swarm learning\cite{warnat2021swarm}, a decentralized machine learning framework that combines edge computing, blockchain based peer-peer networks and coordination, can keep confidentiality without the need for a central server. Therefore, it is also a promising way to implement decentralized recommendation systems.

\textbf{Incentive mechanisms in FedRS.}
FedRS collaborate with multiple participants to train a global recommendation model, and the recommendation performance of the global model is highly dependent on the quantity and quality of data provided by the participants. Therefore, it is significant to design an appropriate incentive mechanism to inspire participants to contribute their own data and participate in collaborative training, especially in the cross-organization federated recommendation scenarios. The incentive mechanisms must be able to measure the clients' contribution to the global model fairly and efficiently.

\textbf{Architecture design for FedRS.}
The recommendation systems in industrial scenarios usually consist of the recall layer and ranking layer, which generate recommendation results on the server side. Considering the privacy of users, FedRS must adopt different designs. A feasible solution is local recalling and ranking, where the server sends the entire set of candidate items to clients, and clients generate recommendation results locally. However, such design brings enormous communication, computation and memory costs for clients since there are usually millions of items in real-world recommendation systems. Another effective approach is to place the recall layer on the server side and the ranking layer on the client side, where clients send encrypted or noised user embedding to the server to recall top-N candidate items, then clients generate personalized recommendation results based one these candidate items via ranking layer\cite{qi-etal-2021-uni-fedrec}. Nevertheless, there is a risk of privacy leakage associated with this approach, because recalled items are known to the server. 

\textbf{Cold start problem in FedRS.} The cold start problem means that recommendation systems cannot generate satisfactory recommendation results for new users with little history interactions\cite{9826710}. In federated settings, the user data is stored locally, so it is more difficult to integrate other auxiliary information (e.g., social relationships) to alleviate the cold start problem. Therefore, it is a challenging and prospective research direction to address the cold start problem while ensuring user privacy.

\textbf{Secure FedRS.}
In the real world, the participants in the FedRS are likely to be untrustworthy. Therefore, participants may upload poisoned intermediate parameters to affect recommendation results or destroy recommendation performance. Although some robust aggregation strategies\cite{Krum} and detection methods\cite{FSAD} have been proposed to defend against poisoning attacks in federated learning settings, most of them don't work well in FedRS. On one hand, some strategies such as Krum, Median and Trimmed-mean degrade the recommendation performance to a certain extent. On the other hand, some novel attacks\cite{FedAttack} use well-designed constraints to mimic the patterns of normal users, extremely increasing the difficulty of detection and defense. Currently, there are still no effective defense methods against these poisoning attacks while maintaining recommendation accuracy.
    
\section{Conclusion}\label{sec:con}
A lot of effort has been devoted to federated recommendation systems. A comprehensive survey is significant and meaningful. This survey summarizes the latest studies on aspects of privacy, security, heterogeneity and communication costs. Based on these aspects, we also make a detailed comparison among the existing designs and solutions. Moreover, we present many prospective research directions to promote development in this field. FedRS will be a promising field with huge potential opportunities, which requires more effort to develop.

\section*{Acknowledgments}
This research is partially supported by the National Key R\&D Program of China 2021YFF0900800, NSFC No.62202279, the Shandong Provincial Key Research and Development Program (Major Scientific and Technological Innovation Project) (No.2021CXGC010108), the Shandong Provincial Natural Science Foundation (No.ZR2022QF018), Shandong Province Outstanding Youth Science Foundation, the Fundamental Research Funds of Shandong University, CCF-Huawei Populus Grove Fund, and the Special Fund for Science and Technology of Guangdong Province under Grant (2021S0053). Sino-Singapore International Joint Research Project (No. 206-A021002).


\bibliography{reference}
\bibliographystyle{IEEEtran}

\vspace{-20pt}
\begin{IEEEbiography}[{\includegraphics[width=1in,height=1.25in,clip,keepaspectratio]{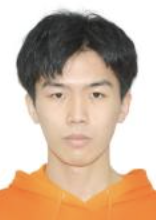}}]{Zehua Sun} is currently pursuing his master's degree in the School of Software of Shandong University. He received his bachelor's degree in software engineering from the School of Software of Shandong University in 2017. His research interests include federated learning, recommendation systems and data mining.
\end{IEEEbiography}
\vspace{-15pt}

\begin{IEEEbiography}[{\includegraphics[width=1in,height=1.25in,clip,keepaspectratio]{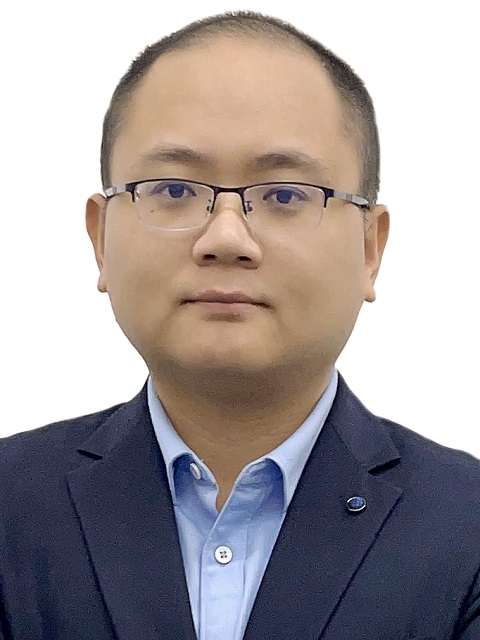}}]{Yonghui Xu}
is a professor at Joint SDU-NTU Centre for Artificial Intelligence Research (C-FAIR), Shandong University, and a research fellow in the Joint NTU-UBC Research Centre of Excellence in Active Living for the Elderly (LILY), Nanyang Technological University, Singapore. He received his Ph.D. from the School of Computer Science and Engineering at South China University of Technology in 2017 and BS from the Department of Mathematics and Information Science Engineering at Henan University of China in 2011. His research areas include various topics in Trustworthy AI, knowledge graphs, expert systems and their applications in e-commerce and healthcare. He has been invited as reviewer of top journals and leading international conferences, such as, TKDE, TNNLS, IEEE Transactions on Cybernetics, Knowledge-Based System, TKDD, IJCAI and AAAI.
\end{IEEEbiography}
\vspace{-10pt}

\begin{IEEEbiography}[{\includegraphics[width=1in,height=1.25in,clip,keepaspectratio]{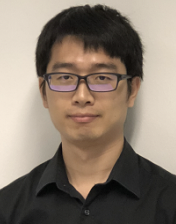}}]{Yong Liu} is a Senior Research Scientist at Alibaba-NTU Singapore Joint Research Institute, Nanyang Technological University (NTU). He was a Data Scientist at NTUC Enterprise, and a Research Scientist at Institute for Infocomm Research (I2R), A*STAR, Singapore. He received his Ph.D. degree in Computer Engineering from NTU in 2016 and B.S. degree in Electronic Science and Technology from University of Science and Technology of China (USTC) in 2008. His research interests include recommendation systems, natural language processing, and knowledge graph. He has been invited as a PC member of major conferences such as KDD, SIGIR, ACL, IJCAI, AAAI, and reviewer for IEEE/ACM transactions.
\end{IEEEbiography}
\vspace{-10pt}

\begin{IEEEbiography}[{\includegraphics[width=1in,height=1.25in,clip,keepaspectratio]{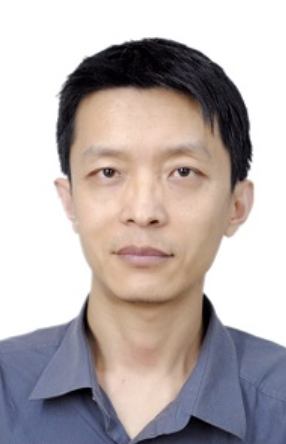}}]{Wei He} is a associate professor at Shandong university. He received bachelor and master degrees from computer science department of shandong university in 1994 and 1999 respectively, and received Ph.d. from engineering of shandong university in 2009. He won the progress first prize in science and technology of shandong province and the progress second prize in science and technology of shandong province, and excellent achievement in computer application. He has published more than 20 papers in the computer journal, journal of software of domestic and international journals conference. More papers were recorded by SCI, EI.
\end{IEEEbiography}
\vspace{-10pt}

\begin{IEEEbiography}[{\includegraphics[width=1in,height=1.25in,clip,keepaspectratio]{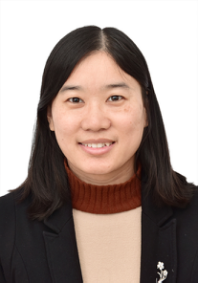}}]{Lanju Kong} is an Associate Professor at Shandong University in Jinan China. she received her 
bachelor degree,  master degree and PH.D from Shandong university in 1999,2002 and 2011respectively. In 2015,she worked in UCSB as visiting scholar for one year.Her research interests include blockchain consensus, multi-chain architecture, large - scale data management, and so on(klj@sdu.edu.cn).

\end{IEEEbiography}
\vspace{-15pt}

\begin{IEEEbiography}[{\includegraphics[width=1in,height=1.25in,clip,keepaspectratio]{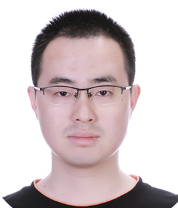}}]{Fangzhao Wu} is a Principal Researcher at Microsoft Research Asia, President of AAAI2022 and senior member of China Computer Society. He received the Ph.D. and B.S. degrees both from Electronic Engineering Department of Tsinghua University in 2017 and 2012 respectively. He published more than 100 academic papers and was cited nearly 3000 times He has won NLPCC2019 Excellent Paper Award, WSDM 2019 Outstanding PC and AAAI 2021 Best SPC. His research mainly focuses on responsible AI, privacy protection, natural language processing, and recommender systems. The research results have been applied in Microsoft News, Bing Ads and other Microsoft products.
\end{IEEEbiography}
\vspace{-15pt}

\begin{IEEEbiography}[{\includegraphics[width=1in,height=1.25in,clip,keepaspectratio]{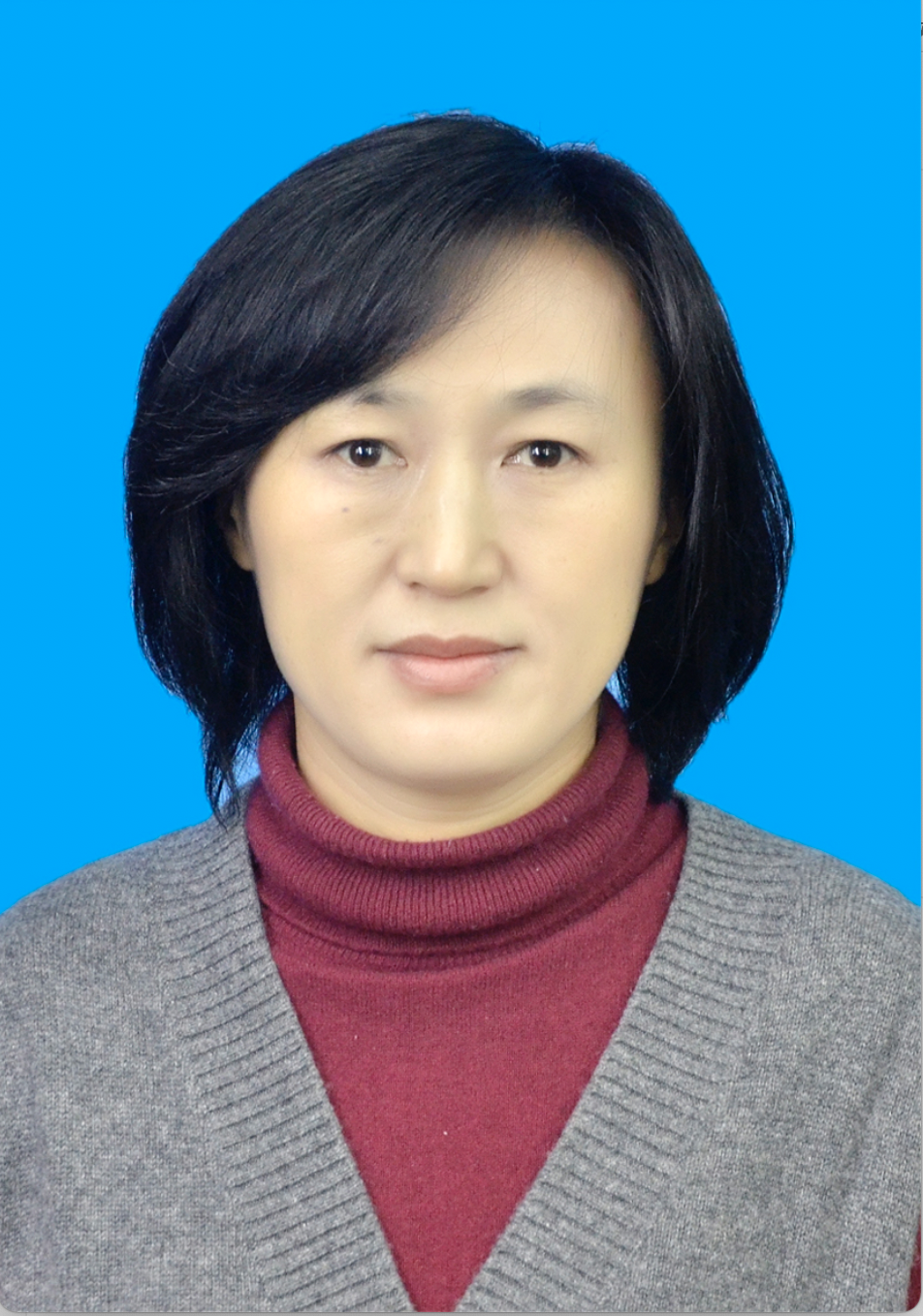}}]{Yali Jiang} is currently a Lecturer in the School of Software, Shandong University. She received her B.Sc., M.Sc. and Ph.D. degrees from Shandong University in 1999, 2002 and 2011, respectively. She is engaged in information security and cryptography research, her main research areas are public key security authentication system and lattice based cryptographic algorithm design and analysis, including cloud computing security, big data privacy protection, IoT security, etc. She has participated in the National 863 Program, Shandong Provincial Excellent Young and Middle-aged Research Award Fund, Shandong Provincial Natural Science Foundation and joint research projects of enterprises.
\end{IEEEbiography}
\vspace{-15pt}

\begin{IEEEbiography}[{\includegraphics[width=1in,height=1.25in,clip,keepaspectratio]{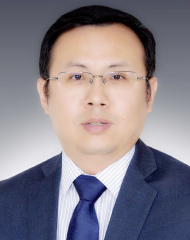}}]{LiZhen Cui} (IET Fellow, IEEE Senior Member) is the Dean at School of Software, Shandong University. He is the Co-Director of Joint SDU-NTU Centre for Artificial Intelligence Research (C-FAIR) and Research Center of Software \& Data Engineering, Shandong University. He is the Associate Director of National Engineering Laboratory for E-Commerce Technologies. He is a Professor with the School of Software and the Joint SDU-NTU Centre for Artificial Intelligence Research (C-FAIR), Shandong University, and also a Visiting Professor with Nanyang Technological University, Singapore. He was a Visiting Scholar with Georgia Tech, Atlanta, GA, USA. He received his bachelor’s, M.Sc., and Ph.D. degrees from Shandong University, Jinan, China, in 1999, 2002 and 2005, respectively. He has authored or coauthored over 200 articles in journals and refereed conference proceedings. His research interests include big data management and analysis and AI theory and application.
\end{IEEEbiography}

\end{document}